\documentstyle[prd,aps,floats,twocolumn,epsfig]{revtex}
\def\be{\begin{equation}}
\def\ee{\end{equation}}
\def\beq{\begin{equation}}
\def\eeq{\end{equation}}
\def\bea{\begin{eqnarray}}
\def\eea{\end{eqnarray}}
\def\p0{\phi_{0}}

\def\hc{\displaystyle{\frac{\dot{a}}{a}}}
\def\Sun{\odot}
\def\bh{bh}

\begin{document}

\twocolumn[\hsize\textwidth\columnwidth\hsize\csname@twocolumnfalse\endcsname
\date{\today}
\title{Could supermassive black holes be quintessential
primordial black holes? }
\author{Rachel Bean and Jo\~ao Magueijo }
\address{}
\maketitle
\begin{abstract}
There is growing observational evidence for a population of
supermassive black holes (SMBHs) in galactic bulges. We examine in
detail the conditions under which these black holes must have
originated from primordial black holes (PBHs). We consider the merging and accretion
history experienced by SMBHs to find that, whereas it is possible
that they were formed by purely astrophysical processes, this is
unlikely and most probably a populations of primordial progenitors
is necessary. We identify the mass distribution and comoving
density of this population and then propose a cosmological
scenario producing PBHs with the right properties. Although this
is not essential we consider PBHs produced at the end of a period
of inflation with a blue spectrum of fluctuations. We constrain
the value of the spectral tilt in order to obtain the required PBH
comoving density. We then assume that PBHs grow by accreting
quintessence showing that their mass scales like the horizon mass
while the quintessence field itself is scaling. We find that if
scaling is broken just before nucleosynthesis (as is the case with
some attractive non-minimally coupled models) we obtain the
appropriate PBH mass distribution. Hawking evaporation is
negligible in most cases, but we also discuss situations in which
the interplay of accretion and evaporation is relevant.
\end{abstract}


]

\renewcommand{\thefootnote}{\arabic{footnote}}
\setcounter{footnote}{0}

\section{Introduction}
There is growing evidence for the presence of supermassive black
holes (SMBHs) in the center of most galaxies
\cite{rich,rich1,rich2,mag,mer,van,barg} including our own
\cite{mel} (but see \cite{mof} for a more skeptical view). The
origin of these black holes (and their relation to the host
galaxy) is far from certain but several theories have been
advanced. Of relevance is the the observational fact that there is
a proportionality relation between the mass of the nuclear
 black hole, $M_{\bh}$, and that of the bulge\cite{mag,mer}. The
bulge mass appears to be about a thousand times larger the
 black hole mass, a relation which holds over 3 to 4 orders of
magnitude.

Nevertheless this close relationship may be interpreted variously,
and ultimately one is confronted with a chicken and egg problem -
what came first: the host galaxies or the SMBHs? It is not
inconceivable that SMBHs are purely the result of the internal
galactic dynamics and their merging history; yet no one has
proposed a concrete mechanism for converting stellar mass objects
into objects 6 to 10 orders of magnitude larger (with the possible
exception of \cite{ostriker}). But it could also be that central
black holes preceded any luminous activity and that  black holes
led the formation of the first galactic bulges and Quasar (QSO)
activity \cite{rich1,rich2,silk}. If the latter is true one must
then find an explanation for the origin of the primordial
population of black holes.

In both scenarios it is inescapable that  black hole growth has
taken place in the recent cosmic history. Even if the budge matter
is well virialized, and whether or not it fuels the SMBH, every
time galaxies merge and their nuclear black holes coalesce part of
the bulge matter ends up in the central black holes \cite{haen}.
Thus, as galactic merging proceeds, the comoving density of SMBHs
decreases and their masses increase. In Section~\ref{mergers} we
spell out the uncertainties of this chicken and egg process,
identifying the conditions under which a primordial population of
black holes is necessary. We then compute the density and mass
profile required of this population, in order to explain the
observed SMBHs.

The rest of our paper is devoted to proposing a cosmological
mechanics for producing the required pre-galactic black holes.
According to our theory SMBHs are descendants of primordial black
holes (PBHs) produced in the very early universe. PBHs are
produced, for example, at the end of inflation\cite{green}, in
double inflation scenarios \cite{kanazawa}, or in first order
phase transitions \cite{jed}. To fix ideas, and although this is
not essential for our paper, we shall follow \cite{green} and
assume that PBHs are produced at the end of a period of inflation
with a blue spectrum of fluctuations (with the possibility of a
running spectral index $n_s$). Whatever their origin, all PBHs
previously considered in the literature are much lighter than
SMBHs, with masses of the order $M\approx M_\Sun (T/1{\rm
Gev})^{-2}$ for PBHs formed at temperature $T$. Hence only a very
unrealistic phase transition, at $T\approx 1{\rm Mev}$ could
produce SMBHs with masses of the right order of magnitude.

However the standard argument assumes that for all relevant cases
PBH masses remain constant once they're formed, since evaporation
and accretion can be neglected~\cite{hawk,bet,cust}. We show that
this is not necessarily true and focus on scenarios in which
significant accretion occurs during the lifetime of PBHs.
Specifically we assume that the universe is pervaded by a
quintessence field~\cite{wett,peeb,fr,quint,andy}. Black holes
cannot support static scalar fields in their vicinity and will try
to ``eat'' them; quintessence is no exception. In the process
their mass increases, so that the seeds which led to the SMBHs we
observe today could be PBHs which have eaten too much
quintessence.

We examine this possibility in Sections~\ref{evac} to~\ref{spect}. In Section~\ref{evac} we estimate the effects of
evaporation and accretion in the presence of a quintessence field. In Section~\ref{comov} we compute the comoving
density of PBHs in our model. Finally in Sec.~\ref{spect} we compute the PBH mass spectrum, and constrain the free
parameters in our model in order to fit the requirement derived in Section~\ref{mergers}.

\section{Galactic black hole accretion and merger histories}
\label{mergers}

As outlined above, the close relationship between the mass of the
central galactic  black hole $M_{bh}$ and that of the galactic
bulge may be interpreted variously. It could signal that the
central  black hole was formed by the inflow of bulge matter
(stars and gas), but it could also be that the central  black hole
was there initially \cite{rich1,rich2,silk} and led the formation
of galactic bulges.
Most probably there was a combination of the two scenarios, and
the central  black hole could indeed be primordial, but at the
same time it was also fed by outside matter.

In this section we show that such a combination is indeed highly
feasible. We devolve the merger history of the galactic halos
\cite{lacey,somerville}, and use a simple merging/ accretion
prescription for the behaviour of the central black holes
\cite{chr}. We thereby show that the majority of galactic black
holes today could have originated prior to the formation of
significant halos, whilst still being in agreement with the
observed mass evolution seen in QSO's \cite{chokshi}.

To devolve the galactic black holes, we use a merger tree approach
\cite{lacey,somerville} to establish the
 histories of the halos and the supermassive black holes within them. The method involves
prescribing a number density distribution of halos nowadays using the Press-Schecter formula \cite{press},
\begin{eqnarray}N(M,z)dM =&& \left({1\over 2\pi}\right)^{1\over
2}\times \nonumber \\\left({\rho\over M}\right)\left({\omega\over \sigma}\right) \left|{d ln \sigma\over d
lnM}\right|&&\exp\left\{-{\omega^{2}\over 2\sigma^{2}}\right\}dM\label{rbeq1}.
\end{eqnarray}
An adaptation of the basic equation provides the conditional
probability that a halo of mass $M_{0}$ at redshift $z_{0}$
evolved from a progenitor of mass $M_{1}=M_{0}-\Delta M$ at
redshift $z_{1}=z_{0}+\Delta z$ \cite{lacey},
\begin{eqnarray}P(M_{0},z_{0}| M_{1}, z_{1})=&&\left({1\over 2\pi}\right)^{1\over 2}\times\nonumber \\
{\omega(z_{0})-\omega(z_{1})\over[\sigma^{2}(M_{1})-\sigma^{2}(M_{0})]^{3\over
2}}&&\exp\left\{-{[\omega(z_{0})-\omega(z_{1})]^{2}\over 2[\sigma^{2}(M_{1})-\sigma^{2}(M_{0})]}\right\}\label{rbeq2}.
\end{eqnarray} We use the notation of Lacey and Cole where in equations (\ref{rbeq1}) and (\ref{rbeq2}), $\rho(z)$
is the energy density, $\sigma^{2}(M)$ is the variance of density
fluctuations on a spherical scale enclosing a mass $M$, and
$\omega(z)=\delta_{c}(1+z)$ where $\delta_{c}\sim 1.686$ is the
over density threshold for density fluctuations to collapse. We
take $\omega(z)$ from a modelled matter power spectrum using
$H_{0}=75 km s^{-1} Mpc^{-1}$ and
$\Omega_{c}=0.25$,$\Omega_{b}=0.05$,and $\Omega_{\phi}=0.7$. The
redshift step $\Delta z$ is mass dependent and represents a
realistic merging timescale for the chosen halo, we follow
\cite{somerville} and take $\Delta z=\Delta\omega(M)/\delta_{c}-1$
where \begin{eqnarray}\Delta\omega(M)\sim\left({\sigma^{2}(M)\over
M} \left|{d ln \sigma^{2}(M)\over d ln
M}\right|\right)^{1/2}.\end{eqnarray}

As a halo is deconstructed only those with a mass greater than a
limiting mass $M_{l}$ are traced, any progenitor with $M<M_{l}$ is
treated as unbound matter accreted onto the halo. We consider a
range of limiting mass scales $M_{l}=10^{9}-10^{10}M_{\odot}$.
This is akin to putting in a lower threshold on the halo's
velocity dispersion $\sigma_{*}\sim v_{vir}=(GM/r_{vir})^{1/2}$ of
$\sim 40-50km s^{-1}$ where $r_{vir}$ is the virial radius and
where we assume a spherical bulge for the halo so that $M=4\pi
r_{vir}^{3}\rho_{halo}/3$. We initially restrict our analysis to
typical galaxy mass scales today $M\sim 10^{10}-10^{12}M_{\odot}$.

We assume that a central  black hole is present in all halos
$M>M_{l}$. To assign  black hole masses in each halo today, we use
the powerful correlation recently found between the  black hole
mass, $M_{bh}$, at the center of galaxies and the line of sight
velocity dispersion $\sigma_{*}$ \cite{gebhardt,sarzi}.
Specifically we use the best fit relation, $M_{bh}=1.2\times
10^{8}M_{\odot}(\sigma_{*}/200 km s^{-1})^{3.75}$\cite{gebhardt}.

The evolution of black holes through merging events and accretion
of halo matter is a complex one, the timescales over which merging
occurs will be intricately dependent on, amongst others, the size
of halos and the ferocity of the merging event. While the
mechanism for accretion will be dependent upon the halos
properties (for example redshift and halo velocity dispersion). We
do not endeavour to model these complex processes here and instead
take a simplistic approach focused on current observational
constraints. We assume halo mergers are violent events allowing
black holes to merge on timescales significantly less than the
time between halo mergers. We then model accretion using a
relation proposed by \cite{chr} whereby a fraction of the halo gas
is accreted onto the  black hole,
\begin{eqnarray}M_{acc}=\epsilon_{acc}(1+z)^{2}M_{halo}\exp[-(\sigma_{*}/300
km s^{-1})^{4}]\label{acceqn}\end{eqnarray} where $\epsilon_{acc}$ is the accretion efficiency factor, and the
velocity dispersion dependence is introduced to account for the reduction in the halo gas's ability to cool in the
gravitational well around the centre at late times.

We are interested in the prospect that primordial black holes are
progenitors for the SMBHs nowadays. We assume that only halos with
$M>M_{l}$ contribute significantly to accretion onto the central
black hole. Subsequently, if a halo only has progenitors of
$M<M_{l}$ and retains a  black hole then the  black hole is
assumed to have not undergone any gas accretion at higher
redshifts and is treated as primordial.

\begin{figure}
\centerline{\psfig{file=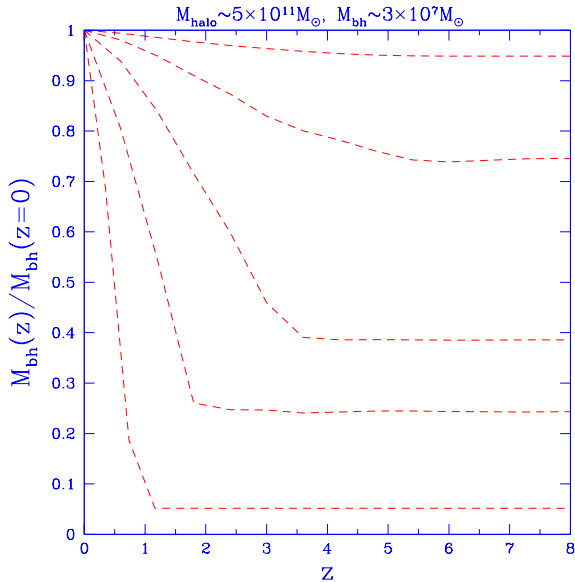,width=8 cm,angle=0}}
\caption{Evolution to give the present day  black hole in a
$5\times 10^{11}M_{\odot}$ halo with redshift for various
accretion efficiency factors. From top to bottom the evolutions
are for $\epsilon_{acc}=10^{-7}, 3\times 10^{-7}, 10^{-6}, 3\times
10^{-6}$,and $10^{-5}$.} \label{eacc1}
\end{figure}

\begin{figure}
\centerline{\psfig{file=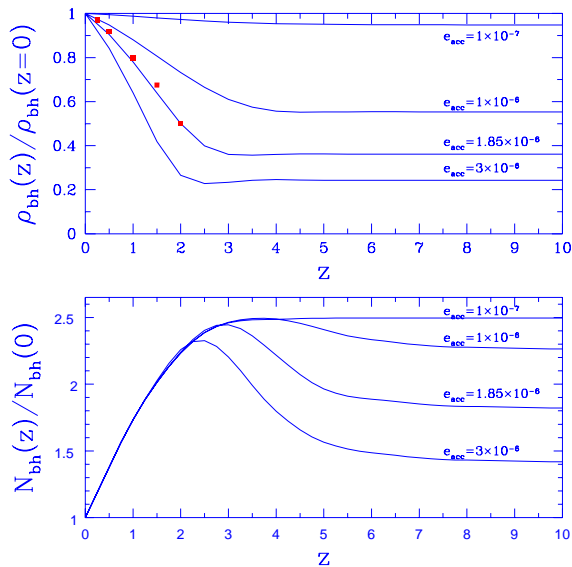,width=8 cm,angle=0}}
\caption{Evolution of mass and number density for galactic black
holes for 4 different accretion efficiencies. The data points are
the inferred mass density values of Chokshi and Turner. The
initial mass density and number density of black holes are taken
to be $1 \times 10^{4} M_{\odot}Mpc^{-3}$ and $4\times
10^{-3}Mpc^{-3}$ at $z=0$ respectively. The number density at
early time reaches $9\times 10^{-3}Mpc^{-3}$ at $z=10$ for the
scenario that is consistent with observations. As one views the
number density evolution from high z, the increase in number
density arises from the formation of astrophysical black holes due
to halo matter accretion. The contribution of astrophysical black
holes becomes more prevalent as the accretion efficiency is
increased.} \label{eacc2}
\end{figure}

Figure \ref{eacc1} demonstrates, for the case of a single halo,
how the accretion efficiency, $\epsilon_{acc}$, in (\ref{acceqn})
has a strong influence on how much of the  black hole mass could
be present at higher redshifts. In figure \ref{eacc2} we show the
evolution of the total comoving energy and number densities for
black holes with redshift. There is an intuitive play off, shown
in figure \ref{eacc2}, between the accretion efficiency and the
fraction of black holes originating at early times. The higher the
accretion rate, the higher the chance is of black holes being
formed at late times, during the halo merging, as opposed to being
primordial. As discussed in \cite{chr}, constraints can be placed
on the efficiency coefficient using inferred accretion rates from
QSO luminosity functions \cite{chokshi}.  One can see that the QSO
evolution data places a tight constraint on the accretion
efficiency parameter: for $M_{l}=10^{10}M_{\odot}$ one requires
$\epsilon_{acc}\sim 1.85\times 10^{-6}$. In this scenario we see
that the vast majority of black holes, using this accretion
prescription, are present prior to halo merging.

\begin{figure}
\centerline{\psfig{file=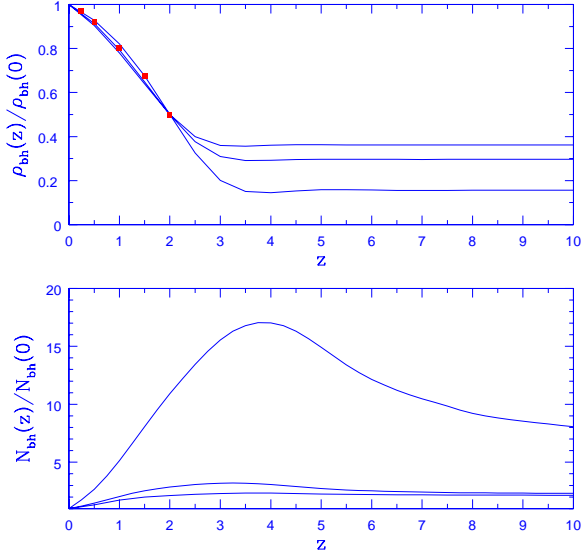,width=8 cm,angle=0}}
\caption{Evolution  of mass and number density of black holes in
galaxy scale halos for 3 different limiting mass thresholds.
Reading top to bottom for mass density and bottom to top for
number density $M_{l}=10^{10},5\times 10^{9}$ and
$10^{9}M_{\odot}$ with accretion efficiencies of
$\epsilon_{acc}=1.85,1.63$ and $1.38\times 10^{-6}$ respectively
chosen to be in agreement with low z QSO observations}
\label{eacc3}
\end{figure}

An alternative scenario extends the galaxies able to support black
holes down to dwarf scales. This can be modeled by lowering the
mass threshold to $10^{9}M_{\odot}$. Inevitably this will alter
the evolutionary profile; accretion in smaller galaxies will be
comparatively more efficient under this model as their
gravitational well is not as prohibitive. As is shown in figure
\ref{eacc3} this leads to a slight reduction in the predicted
comoving energy density of black holes at early times. However,
the general conclusions are the same in that the PBHs still
contribute a significant fraction of the energy density today and
comprise the majority of the total number of black holes involved
in the evolution.

\begin{figure}

\centerline{\psfig{file=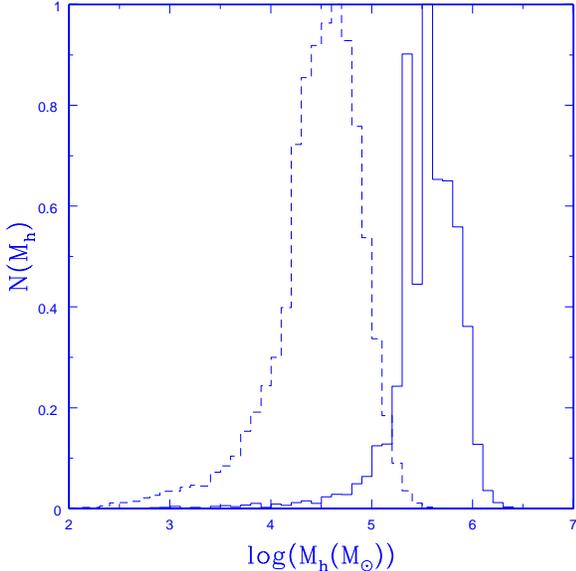,width=8 cm,angle=0}}
\caption{Distribution of  black hole masses at redshifts preceding
halo merger activity for the models in fig 3 with
$M_{l}=10^{10}M_{\odot}$ (full line) and $10^{9}M_{\odot}$ (dashed
line). The distributions are normalised to their peak values.}
\label{eacc4}
\end{figure}

In figure \ref{eacc4} we provide the main output of this section.
We show the expected mass distribution for the primordial
 black hole for $M_{l}=10^{10}M_{\odot}$ and
$M_{l}=10^{9}M_{\odot}$. Their masses, typically $M _{bh}\sim
10^{3}-10^{6}M_{\odot}$, are still many orders of magnitude larger
than stellar mass scales, even considering the potential merging
of early massive stars. We propose that such  black hole masses
could be generated through early time accretion of a
quintessential scalar field onto PBHs. In the following sections
we investigate such a scenario and show that mass distributions
such as those is figure \ref{eacc4} can be obtained.

\section{The effects of evaporation and accretion on primordial black holes}\label{evac}

There has been intense discourse regarding whether or not PBHs are
capable of accreting radiation. Carr and Hawking gave arguments
for negligible accretion \cite{hawk}, but these were later
disputed (see \cite{bet} and references therein). In any case
these arguments only apply to perfect fluids and a scalar field
$\phi$ is not a perfect fluid (even though an isotropic,
homogeneous scalar field does behaves like a perfect fluid). In
Appendix 2 we show that even for the simplest potentials $V(\phi)$
the field is indeed absorbed by the  black hole leading to a mass
increase with rate:
\begin{equation}\label{dm}
{dM\over dt}=4\pi(2M)^2\dot \phi_c^2
\end{equation}
where $\phi_c$ is the cosmological solution for $\phi$. For more
general potentials it may well be the case that the
proportionality constant in the equation receives corrections of
order 1, but these shall not matter for the rest of the argument
in this paper.

From eqn.~\ref{dm} we note that kinetic (but not potential) scalar
energy leads to  black hole growth. This is consistent with the
result that the presence of a cosmological constant does not lead
to equivalent growth.

Assuming now a potential of the form $V=V_0 e^{-\lambda\phi}$, we
have
\begin{equation}
\phi={2\over \lambda \sqrt {8\pi}}\log{t\over t_0}
\end{equation}
(using $G=1$, not $8\pi G=1$), leading to:
\begin{eqnarray}\label{growth}
{dM\over dt}&=&\kappa {M^2\over t^2}\\
\kappa&=&{8\over \lambda^2}
\end{eqnarray}
Eqn.~\ref{growth} integrates to
\begin{equation}\label{acrmass}
{1\over M}= {1\over M_0} + \kappa{\left( {1\over t}- {1\over
t_0}\right)}
\end{equation}
leading to the asymptotic mass:
\begin{equation}
M_\infty={M_0\over 1-{\kappa M_0\over t_0}}
\end{equation}
for black holes with a mass smaller than a critical mass
$M_{crit}=t/\kappa$. These black holes eventually stop accreting
and therefore are subdominant with respect to all others. For
$M=M_{crit}$ BHs grow like $t$.
Above this value the black holes would seem to grow faster that
$t$, however clearly the approximations used must break down at
this stage. In \cite{bet} it was shown that causality constrains
these to grow like $t$ as well, a result we shall use for the rest
of our calculations.

Hence there is a critical mass at time of formation when PBHs may
grow proportionally to the horizon mass. This critical mass
separates those black holes which will be of relevance for out
scenario and those which won't.

\begin{figure}
\centerline{\psfig{file=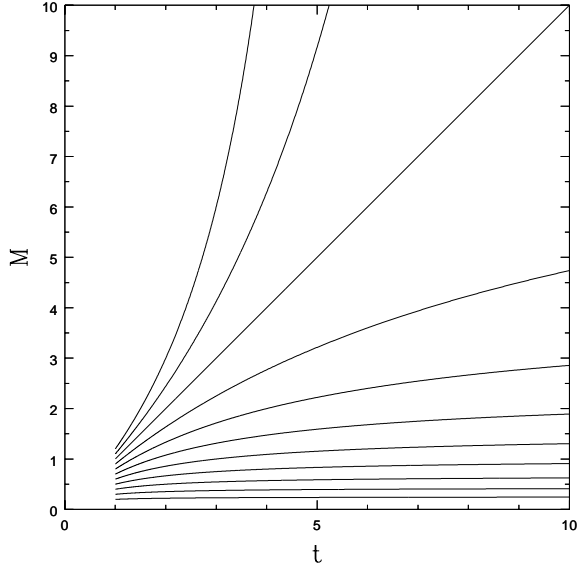,width=8 cm,angle=0}} \caption{The
evolution of black holes ignoring evaporation.   } \label{growthf}
\end{figure}

In addition PBHs may experience significant evaporation, via
Hawking radiation. This leads to a decrease in their mass, at
rate:
\begin{eqnarray}\label{evap}
{dM\over dt}&=&-{\alpha\over M^2}\\
\alpha&=&{\Gamma \over 15360 \pi}
\end{eqnarray}
This equation can be integrated into
\begin{equation}\label{evapmass}
M=(M_0-3\alpha(t-t_0))^{1/3}
\end{equation}
implying an evaporation time of
\begin{equation}
\tau={M_0^3\over 3\alpha}
\end{equation}
or
\begin{equation}
{\tau\over 10^{17} sec}\approx {\left(M\over 10^{15} g\right)}^3
\end{equation}
For black holes formed at temperatures much smaller than the
Planck temperature this effect can be ignored.

Considering now both accretion and evaporation the  black hole
mass rate equation becomes
\begin{equation}\label{acevapeqn}
{dM\over dt}=-{\alpha\over M^2}+\kappa {M^2\over t^2}
\end{equation}
which does not have analytical solutions. However, if the
temperature at which the black holes are produced is not close to
the Planck temperature the interplay of accretion and evaporation
is very simple. Black holes with $M_0>M_{crit}$ will grow with
$M\propto t$, and since their mass was never too small they never
experience significant evaporation. Black holes with
$M_0<M_{crit}$ will stop growing at some $M_\infty$, following
(\ref{acrmass}). If this is smaller than $10^{15}g$ they will
evaporate before today following (\ref{evapmass}). Accretion and
evaporation happen at very distinct times, so although we do not
have an exact solution to (\ref{acevapeqn}) it is an excellent
approximation simply to glue together back to back (\ref{acrmass})
and (\ref{evapmass}).

\begin{figure}
\centerline{\psfig{file=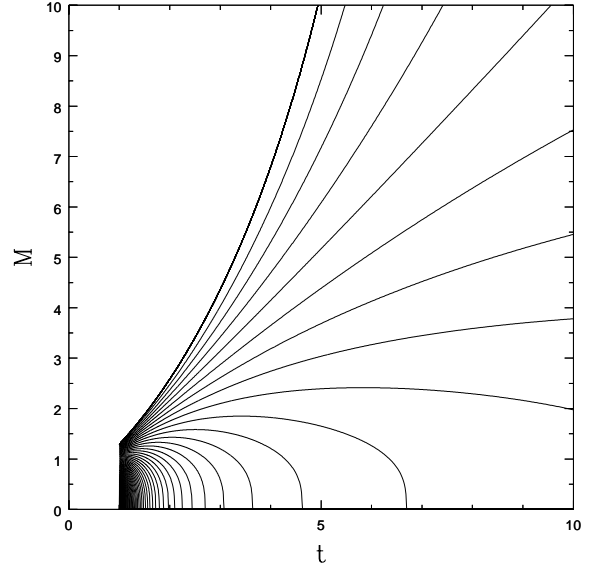,width=8 cm,angle=0}} \caption{The
evolution of black holes formed at Planck time with
$\alpha=\kappa=1$.  } \label{growthev}
\end{figure}

For reheating temperatures of the order of the Planck temperature
the situation is more complicated as evaporation may be
significant while black holes are accreting. The result of a
numerical integration is plotted in Fig.~\ref{growthev} (to be
compared with Fig.~\ref{growthf}). We see that the effect of
evaporation is then to shift upwards the critical mass $M_{crit}$
above which the  black hole mass scales like $t$. In addition,
black holes with masses lower than $M_{crit}$ are quickly removed
by the effects of evaporation even while they are accreting.

\section{PBH formation and their comoving density}\label{comov}

Having identified the conditions required from PBHs in terms of
early-time accretion and late-time merging, we now proceed to
construct a cosmological scenario in which they are satisfied.
Although not compulsory, for definiteness we consider an
inflationary scenario with a reheating temperature $T_r$ and a
tilted spectrum of scalar perturbations with a spectral index
$n_s>1$. The index $n_s$ need not be a constant and indeed many
inflationary models predict a running tilt, varying from scale to
scale. Bearing this in mind we stress that the constraints upon
$n_s$ discussed in this section refer to very small scales (the
horizon size at $T_r$), widely different from the scales probed by
CMB fluctuations or large scale structure surveys.

As shown in \cite{green} for suitable $n_s$ one may have
production of PBHs in a short time window immediately after
reheating. Typically a  black hole is formed if the density
contrast on the horizon scale exceeds a given critical value,
$\delta>\delta_c$. Its mass is given by:
\begin{equation}\label{pbhmass}
M_{bh}=kM_{H}(\delta-\delta_{c})^{1/\gamma}.
\end{equation}
Numerical studies with PBHs formed in a pure radiation background
have identified $\delta_c$, $k$ and $\gamma$ \cite{nim,nim1};
it may be that quintessence modifies these values slightly however
for simplicity we take the calculated values $k=3.3, \gamma=0.34$.
But given the uncertainty
we consider two values of $\delta_{c}$ in this section
$\delta_{c}=0.25,1$. The correct value should be somewhere in
between. In section \ref{spect} we assume the value
$\delta_{c}=0.67$ to outline a precise example.

Note that the horizon scale is important because it is the
relevant Jeans scale for radiation but also for the quintessence
field. In Appendix 1 we present a simple model for the formation
of PBHs with quintessence: some peculiarities are found, but they
should not affect the rest of the argument.

The mass variance on the horizon scale at temperature $T$ is
\cite{green}:
\begin{equation}\label{sigma}
\sigma_H=10^{-4}{\left(10^{-17}{\left({T\over  {\rm Mev}}
\right)}^{-2}\right)}^{1-n_s\over 4}
\end{equation}

To a good approximation, as shown in figure \ref{init}, we may
assume that all PBHs are formed immediately after reheating, since
$\sigma_H$ then decreases sharply making black holes rarer.

\begin{figure}
\centerline{\psfig{file=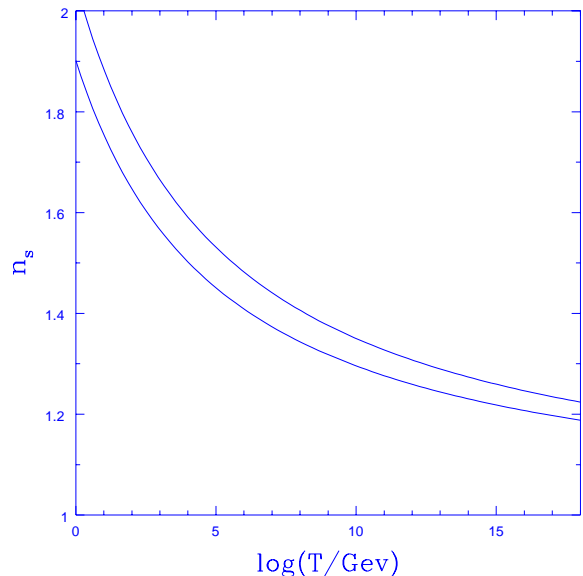,width=8 cm,angle=0}} \caption{The value of the spectral index $n_s$ (on the scale of
the horizon at the time of PBH formation) resulting in the correct comoving density for SMBH seeds. The two curves
correspond to
$\delta_c$=0.25,1 and $\lambda=3$ is taken as an example. }
\label{tilt}
\end{figure}

We have seen that we expect no  black hole coalescence for
redshifts higher than $z\approx 10$. Hence the comoving density of
PBHs can be directly related to the probability of an accreting
 black hole being formed, $p_{acc}$. This is the probability of
 $\delta>\delta_{crit}$, i.e. the
probability for black hole formation and accretion in each
horizon.
\begin{equation}\label{prob}
p_{acc}=\int_{\delta_{crit}}^\infty {e^{-\delta^2\over
2\sigma_H^2}\over {\sqrt 2\pi}\sigma_H}d\delta\approx
{e^{-\delta^2_{crit}\over 2\sigma_H^2}\over 2\pi}
\end{equation}
(where the last approximation comes from \cite{grad}) and where
$\delta_{crit}$ is the value of $\delta$, in equation
(\ref{pbhmass}), that would create a  black hole of critical mass,
$M_{crit}$. The comoving density of accreting black holes
is $n\approx p_{acc}/({4\over 3}\pi r_H^3)$ where $r_H$
is the comoving horizon radius at $T_r$. The latter is given by
$r_H=2ta_0/a\approx 80 (T/{\rm Mev})^{-1}Pc$ so that
\begin{equation}\label{dens}
n=4.6\times 10^{11}p_{acc} {\left(T\over {\rm Mev}\right)}^{3}
{\rm Mpc}^{-3}
\end{equation}
Combining (\ref{sigma}), (\ref{prob}) and (\ref{dens}), and to be consistent with Fig. \ref{eacc2}, requiring that
$n\approx 9\times 10^{-3} {\rm Mpc}^{-3}$ we obtain a value for $n_s$ for each value of $T_r$. In Fig.~\ref{tilt} we
plot the required $n_s(T)$ for the two values $\delta_c=0.25, 1$, using a fixed value of $M_{crit}(\lambda)$. As
mentioned above, the correct value of $\delta_{c}$ should lie somewhere in this region. In Fig.~\ref{tilt} we plot
the required tilt for three different critical masses with fixed $\delta_{c}=0.67$.

The tilts required fit within the constraints of \cite{liddle},
and can accommodate the recent reionisation tilt constraint of
\cite{he} with $n_{s}<1.27$ with a high reheat temperature $>\sim
10^{15}Gev)$.

\begin{figure}
\centerline{\psfig{file=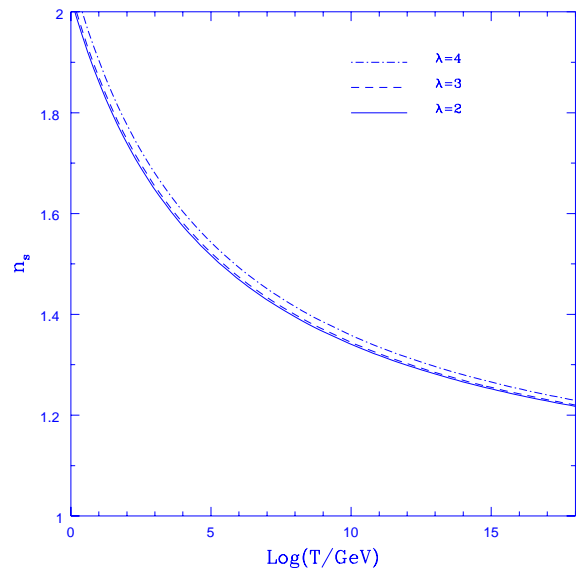,width=8 cm,angle=0}} \caption{The value of the spectral index $n_s$ (on the scale of
the horizon at the time of PBH formation) for three different critical masses relating to $\lambda=2,3,4$. $\delta_c$ is fixed =0.67. } \label{tilt2}
\end{figure}

 Note that the primordial  black hole comoving density imposed in our
considerations is much smaller than that considered in
\cite{green}. Indeed the black holes studied by \cite{green} do
not grow (as opposed to ours
). The idea in \cite{green} is to produce a larger density of much
lighter black holes, suitable to promote them to candidates for
dark matter. Our purpose is to generate a much lower density of
much heavier PBHs, so that they could supply the primordial eggs
for the merging history of SMBHs.

\section{The PBH mass spectrum}\label{spect}

\begin{figure}
\centerline{\psfig{file=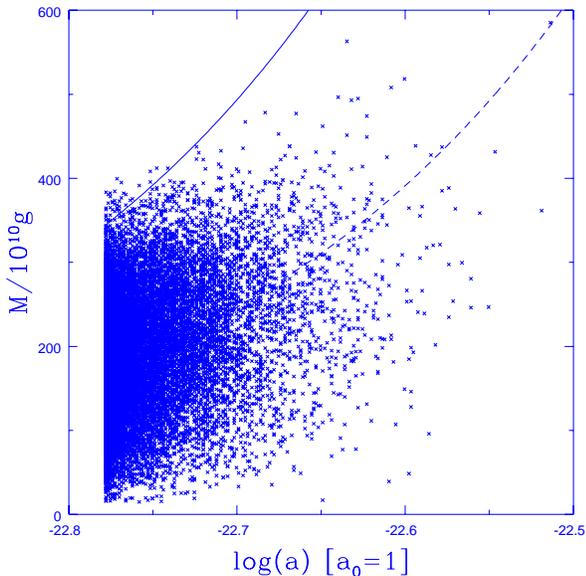,width=8 cm,angle=0}} \caption{The
demography of  black hole formation with a reheat temperature of
$T_r=10^{10}Gev$ and tilt $n_{s}=1.33$. The crosses
represent Monte Carlo simulated black holes. The full and dashed lines are the
horizon and critical mass, the latter assumes $\lambda=2$.} \label{init}
\end{figure}

\begin{figure}
\centerline{\psfig{file=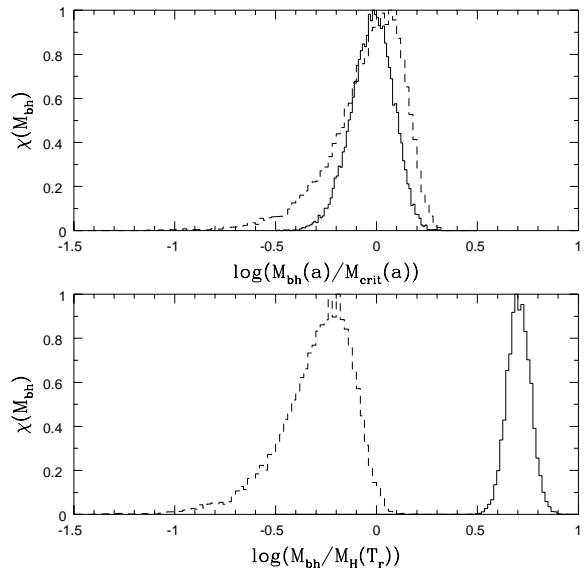,width=8 cm,angle=0}} \caption{The bottom panel shows the initial black hole mass distributions for the two scenarios
described in the text, with reheat temperatures $(T_{r})$ of $10^{10}Gev$
(dashed line) and $10^{4}Gev$ (full line). The upper panel shows the same distributions comparing them to the critical mass at the time of each black hole's formation. A blackhole will accrete if $(M_{bh}/M_{crit})>1$. In each panel the distributions are normalised to 1 at their peak for ease of comparison.} \label{hist1}
\end{figure}

As an example we calculate the mass spectrum following the methods
of \cite{green} to evaluate the initial mass spectrum. This
involves using a Monte-Carlo technique to obtain  black hole
masses which, unfortunately, is computationally unfeasible for low $p_{acc}$ $(\sim <10^{-7})$ required to give the correct comoving densities for $\lambda>2$. With
this in mind, we consider two scenarios which provide the correct
comoving number density of PBHs ($9\times 10^{-3} Mpc^{-3}$
assuming a limiting mass of $10^{10}M_{\odot}$ in section
\ref{mergers}), one with a reheat temperature of the order
$T_r=10^{10}Gev$, tilt $n_{s}=1.33$ and a second
with reheat temperature $T_r=10^{4}Gev$, $n_{s}=1.57$ with
$\lambda=2$ in each case. As Fig.~\ref{init} shows most black holes, and in
particular those that are able to accrete, are then formed
immediately after reheating, with a slight spread in their masses,
which we further display in Fig.~\ref{hist1}.

We then evolve this initial mass distribution considering
accretion and evaporation in a quintessence model as described
in Section~\ref{evac}. We assume, however, that the quintessence field
goes off scaling and kinates at a temperature of the order
$T=1Mev$, since if the field continues to scale after this time one 
ends up with too large masses. Nevertheless if the 
field goes off scaling the black holes stop
accreting (cf. Eqn~\ref{dm}), 
and even if the field starts scaling subsequently, BHs will
not grow significantly again, since their growth pause has rendered
them subcritical. Hence in what follows what we need
is simply a brief {\it pause} in scaling at a temperature of around
$T=1Mev$.  We shall consider two examples here, one with 
$T_{off}\approx 4 Mev$, and another with $T_{off}=1 Mev$.

\begin{figure}
\centerline{\psfig{file=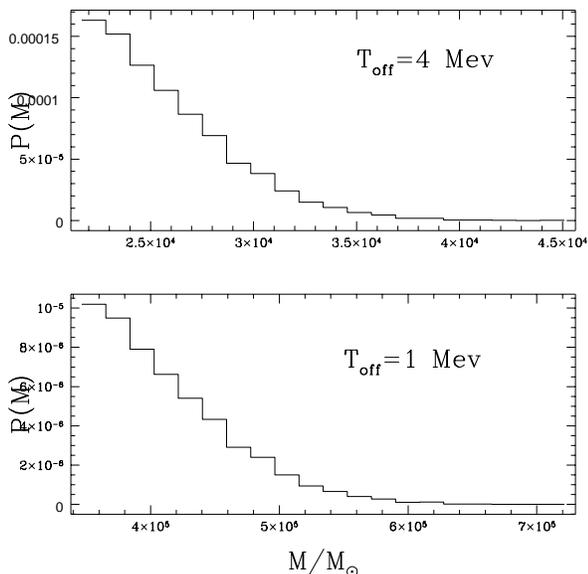,width=8 cm,angle=0}}
\caption{The distribution of black hole masses at redshift
$z\approx 10$ for quintessence models deviating from scaling at
$T\approx 5 Mev$, and $1 Mev$ (we have used a reheating temperature
of $T_r=10^{4}$ Gev and $\lambda=2$). In both cases the distribution is
skewed, reflecting the exitence of a critical mass for growth.
The later quintessence leaves scaling the larger the BHs' mass. } 
\label{hist2}
\end{figure}

\begin{figure}
\centerline{\psfig{file=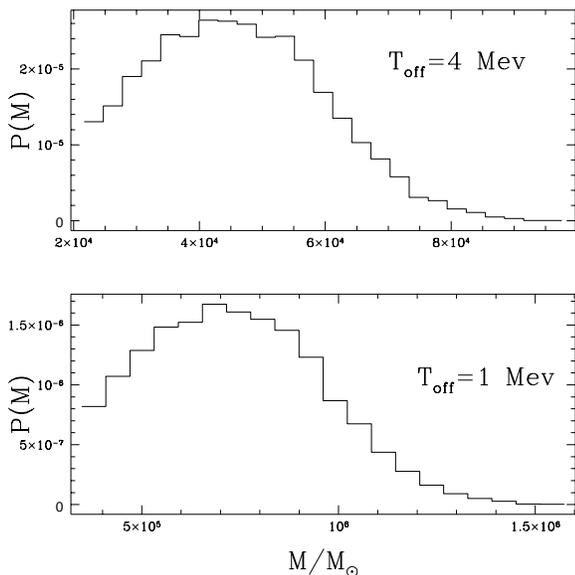,width=8 cm,angle=0}}
\caption{The distribution of black hole masses at redshift
$z\approx 10$ for quintessence models deviating from scaling at
$T\approx 5 Mev$, and $1 Mev$ (we have used a reheating temperature
of $T_r=10^{10}$ Gev and $\lambda=2$). } 
\label{hist3}
\end{figure}

In Figs.~\ref{hist2} and \ref{hist3} we find that the projected mass 
spectra at $z\approx 10$ (or indeed at any time after $T_{off}$
but before the galactic merging history started). In Fig.~\ref{hist2} 
we plot results for PBHs formed at a reheating temprature of
$T_r=10^{4}$ Gev as described above. In Fig.~\ref{hist3} 
we plot the corresponding distribution for the scenarios in which
$T_r=10^{10}$ Gev. In both
cases note that the existence of a critical mass for accretion
implies that the final distribution mimicks the initial one clipped
at the critical mass. For the first scenario considered the cut off
is at the peak of the distribution; hence the final distribution
is very skewed. For the second scenario the cut off is to the left
of the peak - so the final distribution is more symmetric. As expected
if the field goes of scaling later, the final PBH masses are much larger:
for $T_{off}=4$ Mev we find final PBH masses of the order of $10^4-10^5
M_\odot$; for $T_{off}=1$ Mev these masses grow to $10^5-10^6
M_\odot$.

In either case these plots are entirely
consistent with those obtained from the merging history in section
\ref{mergers} (Fig. \ref{eacc4}).
Quintessence could have therefore provided the primordial seeds
which then turn into the SMBHs we observe today.

\section{Conclusions}

We have demonstrated that a scenario in which primordial black
holes attain super massive size through the accretion of a
cosmological scalar field is wholly consistent with current
observational constraints. Such a model can generate the correct
comoving number density and mass distribution for SMBH's, given a
standard prescription for late-time merging and matter accretion
and with reasonable choices of tilt. 

Existing schemes explaining SMBHs require very contrived choices
of parameters (e.g.\cite{ostriker}).
The root of all the evil lies in the huge mass discrepancy between
cosmic scale masses (such as those considered in previous studies
of PBHs) and those of their supermassive cousins - the
cosmological counterpart of the discrepancy between solar mass
objects and SMBHs found in astrophysical schemes. We attempt to
bridge the mass scale gap in our model by allowing primordial
holes to grow by accreting quintessence. Still,  we find
that we have to switch off this process at a carefully tuned time,
so that there is not too much growth.

The fact that this requirement is equivalent to requesting that
quintessence goes off scaling just before nucleosynthesis is of
some consolation: such a feature is already present in
non-minimally coupled quintessence models such as those studied by
\cite{rachjoao}. In these scenarios the present acceleration of
the universe results from a coupling between quintessence and dark
matter. It switches on close to the radiation to matter
transition, but is affected by a long transient. This explains why
the universe did not start accelerating until nowadays. The fact
that this transient is symmetric around equality, and that
equality is an equal number of expansion times from us and from
nucleosynthesis, then makes the field kinate away at
nucleosynthesis time.

In addition, in order to obtain the correct comoving density we
have to tune carefully the value of the scalar tilt, $n_s$, on the
scale of the horizon at the time of primordial  black hole
formation. However this fine tuning is a problem with any theory
employing primordial black holes for astrophysical purposes, such
as as candidates for dark matter \cite{green}, and is no better or
worse in our theory.

In spite of these fine tuning problems we believe that this
is an interesting scenario which deserves further work. The
effects of angular momentum upon the whole picture are perhaps
the most important next issue to consider. 

{\bf Acknowledgements} We would like to thank Andrew Jaffe, Lev Kofman, Andrew
Liddle, David Spergel and Paul Steinhardt for their helpful
thoughts and comments. RB is supported by PPARC.

\section{Appendix I - PBH creation and quintessence}
In this appendix we the examine the effects of quintessence on PBH
formation using the spherical model. The idea is to follow
non-linear collapse of a super-Jeans size spherical region by
modelling the overdense region as a portion of a Friedmann closed
model pasted onto a flat model. Setting up this model entails
studying the dynamics of quintessence in closed models. For
completeness we shall also consider open models, which may be of
relevance for modelling the void structure of out universe.

\subsection{Quintessence in open and closed models}
The relevant equations are:
\begin{eqnarray}
{\left(\dot a \over a\right)} ^2 &=& {1\over 3}{\left(\rho
+\frac{1}{2}\dot{\phi}^{2}+V(\phi)
\right)} -{K\over a^2}\nonumber\\
\rho+3\hc(\rho+p)&=&0 \\
\ddot{\phi}+3\hc\dot{\phi}&=&\lambda e^{-\lambda\phi}
\end{eqnarray}
where dots represent derivatives with respect to proper time. It
is well know that if $K=0$ with $p=w\rho$ we have:
\begin{eqnarray}
a&\propto& t^{2\over 3(1+w)}\\
\rho&\propto& {1\over a^{3(1+w)}}\\
w_\phi&=&w\\
\Omega_\phi&=&{\rho_\phi\over \rho+\rho_\phi}={3\over \lambda^2}(1+w)\\
\phi&=&{2\over \lambda}\log{t\over t_0}
\end{eqnarray}
We call this solution the scaling solution. With $K\ne 0$ this is
also the solution at early times, when curvature is subdominant.

Without quintessence open models ($K=-1$) at late times become
vacuum dominated - the so called Milne Universe for which
$a\propto t$. However we find that in the presence of quintessence
the onset of negative curvature domination leads to another
scaling solution: one in which curvature and quintessence remain
proportional. We find that:
\begin{eqnarray}
a&\propto& t\\
\rho_\phi&\propto& {1\over a^2}\\
w_\phi&=&-{1\over 3}\\
\Omega_\phi&=&{\rho_\phi\over \rho_c}={2\over \lambda^2}\\
\phi&=&{2\over \lambda}\log{t\over t_0}
\end{eqnarray}
which implies that open universes at late times are devoid of
normal matter, but not of quintessence. This type of behavior can
be understood from the $K=0$ scaling solution, considering that
open curvature behaves like a fluid with $w=-1/3$. In the same way
that quintessence locks on to matter or radiation in a scaling
solution, it locks on to open curvature.

As is well known, without quintessence closed models ($K=1$)
expand and eventually turn around and collapse in a Big Crunch.
This type of behavior is not changed by the presence of
quintessence; however the scaling behavior of quintessence itself
is drastically changed. As the universe comes to an halt the
friction term in the $\phi$ equations (that is the term $3{\dot
a\over a}\dot\phi$) is withdrawn and the field becomes kinetic
energy dominated (i.e. it kinates). Hence as the universe turns
around quintessence becomes subdominant, as it scales like
$\rho_\phi\propto 1/a^6$ in contrast with $\rho\propto 1/a^3$.

However as the universe enters the contracting phase what used to
be a friction term starts to drive the field, since now ${\dot
a\over a}<0$. This leading to runaway kination, since it is
precisely the balance of braking and the slope of $V(\phi)$ what
usually moderates the balance of kinetic and potential energy. As
before, kinetic energy domination implies $w_\phi=1$ and a faster
decay rate with expansion $\rho\propto 1/a^6$. However contraction
reverses the argument, and whatever has stronger dilution rate
during expansion, will have higher compression rate during
contraction. Therefore, at late stages of collapse quintessence
dominates. Since curvature and background matter can be ignored we
have the solution:
\begin{eqnarray}
a&\propto& (t_c-t)^{1/3}\\
\rho_\phi&\propto& {1\over a^6}\\
w_\phi&=&1\\
\Omega_\phi&\approx&1\\
\phi&\propto&a^{-1/{\sqrt 6}}
\end{eqnarray}
in which $t_c$ is the crunch time.

\subsection{Implications for structure formation and PBH formation}
Qualitatively these results indicate that quintessence has a
leading role in the strongly non-linear stages of structure
formation. Voids should be filled with quintessence, judging from
what happens to the $K=-1$ case. Also it would appear that
 black hole formation would be led by quintessence and not by matter
(as implied by the $K=1$ case). Interestingly, quintessence
appears to behave like a stiff fluid ($p_\phi=\rho_\phi$) during
collapse, so we should be able to simplify the calculations. The
 black hole mass seems to be dominated by the amount of quintessence
accreted. Also because $M_\phi\propto \rho_\phi a^3\propto 1/a^3$,
there must be a mass enhancement during collapse (due to gravity
acting against quintessence's pressure).

All of these effects may at most introduce factors of order 1 in
the calculations in the main body of the text, and so interesting
as these results might be we have relegated them to Appendix.

\section{Appendix II - Quintessence around a  black hole}
Given the large Jeans mass of quintessence, its fluctuations should
be very small even in the interior of non-linear objects,
such as the Solar system. Indeed the relativistic restoration force
induced by the $\phi$ gradients should ensure that quintessence fluctuations
remain linear, even under highly non-linear gravitational forces.
One obvious exception  is the vicinity of a BHs horizon, where $\phi$
has to change drastically.

The interaction of BHs and quintessence then becomes a problem
of boundary matching: on one side the homogeneous cosmological solution;
on the other the infinite radius limit of the Schwarzchild solution.
This matter has been examined in the literature in the context of
the gravitational memory problem in Brans-Dicke theories
\cite{memory,memorya}. Our analysis will closed mimic that of
Jacobson \cite{jacobson}.

We assume that the cosmological scalar field generates a much
weaker gravitational field than the  black hole, so that we can
impose a Schwarzchild metric, with a quasi-stationary mass
parameter. Hence the equation for the scalar field is the vicinity
of the BH is:
\begin{equation} \label{wave}
\Box\phi=-{\ddot \phi\over A}+
{1\over r^2}(r^2 A \phi')' ={\partial V\over
\partial \phi}
\end{equation}
where $A(r)=1-2M/r$, and dots and dashes are derivatives with
respect to time and $r$ respectively. For free waves ($V=0$) the
system separates giving:
\begin{equation} \phi=e^{-i\omega t}R(r)
\end{equation} with: \begin{equation} {1\over r^2}(r^2 A R')' + {\omega^2 \over A}R=0
\end{equation}
or introducing a Kruskal coordinate $r^\star$, such that
$Adr^\star =dr$ (or $r^\star =r + 2M \log ({r\over 2M} -1 )$):
\begin{equation}
{d^2 R\over dr^{\star 2}} + {2A\over r} {d R\over dr^{\star}}+
\omega^2 R=0
\end{equation}
It is clear that far away the solutions are:
\begin{equation}
\phi={e^{i\omega(t\pm r)}\over r}
\end{equation}
and that near the horizon they become:
\begin{equation}
\phi=e^{i\omega(t\pm r^\star)}
\end{equation}
Focusing on solutions regular
on the horizon ${\cal H}^+$, we introduce the advanced time
coordinate $v=t+r^\star$, so that the relevant oscillatory solutions
take the form $e^{i\omega v}$.

In addition there are non-oscillatory solutions regular at $r=2M$,
such as those studied by Jacobson. These can be obtained noting
that (\ref{wave}) can be rewritten as:
\begin{equation}\label{wavev}
2\phi_{,rv}+{2\over r}\phi_{,v} +{1\over
r^2} (r^2 A\phi_{,r})_{,r}=-V'(\phi)
\end{equation}
For a general $V(\phi)$ separability is lost and numerical work is
necessary. In order to further the analytical approach we model
the rolling potential by its gradient at a point $V=-\mu\phi$
(we could add a constant here without loss of generality).
Equation (\ref{wavev}) then becomes:
\begin{equation}
2\phi_{,rv}+{2\over r}\phi_{,v} +{1\over
r^2} (r^2 A\phi_{,r})_{,r}=-\mu
\end{equation}
Setting $\phi=f(v)+g(r)$ leads to $f=Bv + D$, and
\begin{equation}
g_{,r}={-\mu r^2\over 3(r-2M)} -{Br\over r-2M} + {C\over r(r-2M)}
\end{equation}
For the solution to be regular at the horizon we must have:
\begin{equation}
C=4M^2{\left( B +{2\over 3}\mu M\right)}
\end{equation}
Integrating finally leads to:
\begin{eqnarray}
\phi&=&B[v-r-2M\log r] -\nonumber\\
&&-{\mu\over 3}{\left( {r^2\over 2} + 2M r + 4M^2 \log r\right)}
+D \end{eqnarray}
which generalized Jacobson's solution. The general solution is a
superposition of free oscillatory solutions  and
this solution.

We must now impose a boundary condition of the
form:
\begin{equation}
\phi=\phi_c +\dot \phi_c t
\end{equation}
where $\phi_c$ refers to the quintessence cosmological solution.
If we fix $B$ and $D$ so that:
\begin{eqnarray}
\phi&=&\phi_c + \dot\phi_c
\left[v-r-2M\log {r\over 2M}\right] -\nonumber\\
&&-{\mu\over 3}{\left( {r^2\over 2} + 2M r + 4M^2 \log {r\over
2M}\right)}
\end{eqnarray}
we have asymptotically:
\begin{equation}
 \phi=\phi_c + \dot\phi_c t -{\mu\over 6}r^2
\end{equation}
A small error is made in matching the two conditions, but in a
quintessence scenario with
\begin{equation}
\mu\approx {\partial V\over \partial
\phi}= \lambda V=\lambda \Omega_\phi\rho
\end{equation}
(where $\lambda$ has units of $1/\phi$ and is expressed in Planck units
and  $\rho$ is the cosmological density) we find thatthe scale on which the
error becomes significant is the cosmological horizon scale.
Hence this small error can be neglected.

Computed the flux of energy through the BH horizon associated
with this solution:
\begin{equation}
{\cal F}= T_{vv}=\dot\phi_c^2
\end{equation}
we finally find an equation for the BH mass:
\begin{equation}
{dM\over dt}=4\pi(2M)^2\phi_c^2
\end{equation}
which is the same formula derived by Jacobson. Hence we conclude that
the potential $V(\phi)$ only has indirect impact upon the BH mass,
via its effect upon the time evolution of the cosmological solution
$\phi_c$.

\end{document}